\documentstyle[budapest1]{article}
\RequirePackage{rawfonts}
\frompage{000} \topage{000}                                              
\def\rightmark{ Inflation of fireballs ...}  
\def\leftmark{ T. Cs\"org\H{o} and J. Zim\'anyi}

\def\bea{\begin{eqnarray}}
\def\eea{\end{eqnarray}}
\def\be{\begin{equation}}
\def\ee{\end{equation}}

\def\p{\prime}

\newcommand{\gton}{\mathrel{\lower.5ex \hbox{$\stackrel{> } 
 {\scriptstyle \sim}$}}} 
\newcommand{\lton}{\mathrel{\lower.5ex \hbox{$\stackrel{< } 
 {\scriptstyle \sim}$}}} 
\newcommand{\ben}{\begin{enumerate}} \newcommand{\een}{\end{enumerate}} 
\newcommand{\bit}{\begin{itemize}} \newcommand{\eit}{\end{itemize}} 
\newcommand{\bc}{\begin{center}} \newcommand{\ec}{\end{center}} 
\newcommand{\beqar}{\begin{eqnarray}} \newcommand{\eeqar}[1]{\label{#1} 
\end{eqnarray}}

\title{ 
Inflation of fireballs, the gluon wind and the homogeneity of 
the HBT radii at RHIC
}
\authors{ 
{T. Cs\"org\H{o}$^1$ and J. Zim\'anyi$^1$
}\\[2.812mm] 
{\normalsize 
\hspace*{-8pt}$^1$  MTA KFKI RMKI , H - 1525 Budapest 114, 
  P.O. Box 49,  Hungary 
}}

\abstract{ 
We solve analytically the ellipsoidally expanding fireball hydrodynamics with
source terms in the momentum and energy equations, 
using the non-relativistic approximation.
We find that energy transport from high $p_t$ jets of gluons to 
the medium leads to a transient, exponential inflation of  
the fireballs created in high energy heavy ion collisions.
In this transient, inflatory period, the slopes of the
single particle spectra are exponentially increasing, while
the HBT radius parameters are exponentially decreasing with time.
This effect is shown to be  similar to the development of
 the homogeneity of our Universe due to an inflatory period.
Independently of the initial conditions, 
and the exact value of freeze-out time and temperature,
the measurables (single particle spectra, 
the correlation functions,
slope parameters, elliptic flow, HBT radii and cross terms) 
become time-independent during the
late, non-inflatory stages of the expansion, 
and they satisfy new kind of scaling laws.
If the expansion starts with a transient inflation caused by the
gluon wind, it leads naturally to large transverse flows as
well as to the the simultaneous equality,
and scaling behaviour of the HBT radius parameters, 
$R_{side}\approx R_{out} \approx  R_{long} \approx t_f \sqrt{T_f/m}$.
With certain relativistic corrections, the scaling limit is 
$\tau_f \sqrt{T_f/m_t}$, where $m_t$ is the mean transverse mass of the pair.
}

  
\begin{document} 
  
\maketitle 
\setcounter{page}{1} 
 
\section{Introduction}\label{intro} 
Recently, there were a number of QCD based perturbative calculations
that indicated the energy loss of gluon jets with high transverse momentum
due to multiple scattering during the course of their
penetration through hot and dense hadronic matter. This phenomena,
the jet quenching goes back to papers of Gyulassy, 
Pl\"umer and Wang~\cite{Wang:1994fx,Gyulassy:1993hr}, recently studied in 
the papers of Gyulassy, L\'evai  and Vitev (GLV)~\cite{Gyulassy:2000fs}.
It is clear from these works, that the jet quenching mechanism
results in the energy loss or the depletion of the number of hadrons
from the high transverse momentum part of the spectrum. Due to the
conservation of the total energy, this implies that the number of
particles with small total momentum has to increase. See ref.
\cite{Vitev:2002vr} for the summary of recent developments in this field from the
point of view of perturbative QCD.

In this paper, we focus on the phenomenological
consequences of energy and momentum transfer to the soft,
hydrodynamically behaving, small transverse momentum part of the
single particle spectrum. We present a new family of
solutions of non-relativistic hydrodynamics with source terms.
This family of solutions is a generalization of the results of
refs.~\cite{ellobs,ellsol,cssol}
to ellipsoidally symmetric, expanding fireballs that are subject to
energy and momentum transfer from the non-hydrodynamically behaving modes.
A constant rate of energy pumping is shown to lead to
an exponential inflation of the principal axis of the fireballs,
while momentum transfer leads only to linear increase.
These effects are correlated in time to the transition of
high energy gluons through the nuclear medium, hence,
by default, they are also transient phenomena, existing only
in the initial phase of the expansion.
We point out the similarity of this mechanism to the
existence of a transient inflatory period in the expansion
of the early Universe.

Inflation of the Universe after the Big Bang leads to
a homogeneous and flat solution of Einstein's equation.
We show here that an inflation of hadronic fireballs,
caused by the gluon wind in the initial stage of
the Little Bangs of heavy ion collisions
also leads to the simultaneous equality, spherical symmetry
and scaling of the effective, measurable source sizes 
(frequently referred to as the HBT radii). This effect 
connects the physics of the smallest observable scales
to the physics of the largest observable ones,
underlining the universal nature of the laws of physics
and the scale independence of the local conservation laws,
governing the hydrodynamics of these expansions.

\section{There blows the gluon wind ...}\label{inflation}   

Consider the non-relativistic hydrodynamical problem, as given by 
the continuity, Euler and energy equations, appended with source terms
in the momentum and the energy conservation laws: 
\begin{eqnarray} 
{\partial_t n} + {\rm{\bf\nabla}} ({\rm{\bf v}} n) & = & 0\,,
\label{e:cont} \\ 
{\partial_t {\rm{\bf v}}} + ({\rm{\bf v}}{\rm{\bf\nabla}})
{\rm{\bf v}}
& = & - ({\rm {\bf \nabla }} (p+ p_G)) / (m n)\, ,  \label{e:Eu} \\
{\partial_t \epsilon} + {\rm{\bf\nabla}}(\epsilon{\rm{\bf v}}) + 
 p {\rm {\bf \nabla }} {\rm {\bf v }} & = & j_G\, ,  \label{e:en}
\end{eqnarray}
where $n$ denotes the particle number density, ${\rm {\bf v}}$ 
stands for the non-relativistic (NR) flow velocity field, 
$\epsilon$ for the energy density, $p$ for the pressure and in the 
following the temperature field is denoted by $T$. 
The above set of equations are supplemented by the equations of state
(EoS). The parameters of the EoS vary in various
domains of the $(T,\mu)$ plane, corresponding to various phases
of matter. A general, and analytically solvable family of
equations of state is given by 
\begin{eqnarray}
p & = & \lambda_p n T - B\,, \qquad
\epsilon  =  \, \lambda_e n T\, + B.  \label{e:eos}
\end{eqnarray}
The parameters $\kappa = \lambda_e/\lambda_p$  and $B$ characterize the EoS for a broad variety of
materials: e.g. a non-relativistic ideal gas yields  $\kappa=3/2$ and $B=0$.
Softening of the EoS can be modeled by increasing the values of
$\kappa$ in certain temperature domains.
A parameter $B>0$ can be used to describe a
kind of  generalized bag model equation of
state (vacuum pressure) and to describe phase 
transitions, similarly to the class of models
considered in refs.~\cite{rellsol}. 
Hence we present generalizations of the results in refs.
~\cite{ellsol,ellobs} and references therein, for a possible
description of phase transitions during the time evolution of the
exploding fireballs.

The source terms $p_G$ and $j_G$ in eqs.~(\ref{e:Eu},\ref{e:en})
stand for the contribution of 
the gluon wind that blows from the center of the
collision zone through the material for a finite period of time.
This wind does not carry conserved charges,
and we assume that the energy and the momentum transfer is proportional
to the number density $n$ of the fluid, as well as to the average
local velocity of the fluid elements. We assume that the gluons
are proceeding essentially with the speed of light, so the
relative velocity distribution depends only on the breadth of
the local momentum distribution, which in turn is characterized
with the magnitude of the temperature $T$. We define a  model that
is analytically solvable by assuming that
\bea
	p_G & = & j_v(t)\, n T, \\
	j_G & = & j_E(t) \, \kappa n T,
\eea
where the coefficients $j_v$ and $j_E$ depend only on the time variable 
$t$. The time dependence of the gluon wind can be expressed
with the help of the dimensionless functions
$j_v(t)$ and $j_E(t)$
 in such a way that they are non-vanishing only within a finite
period of time $t_1 \le t \le t_2$. The equations of the usual
non-relativistic hydrodynamics
are recovered in the case of vanishing source terms, $j_E(t) = j_v(t) = 0$.

A simple analytic solution of the non relativistic hydrodynamic
equations without source terms (i.e. $j_E = j_v = 0 $) was given
for the spherical and cylindrical symmetric cases in
refs.~\cite{jnr} and ~\cite{dgsbz} more than 20 years ago,
and a recent series of papers~\cite{cspeter,rsol,ellsol,cssol} generalized
these solutions to various (Gaussian, donut shaped or oscillatory) density
and corresponding temperature profiles, for various symmetry classes
ranging from spherical to ellipsoidally symmetric cases. Ref. ~\cite{cspeter}
also considered the case of Gaussian density profiles, spherically symmetric
expansions and source terms in the energy equation, and pointed out the
existence of an inflatory solution with an exponentially increasing radius
parameter for the case corresponding to $j_E(t) = j_0 > 0$. Here we
generalize this solution for the case of ellipsoidal symmetry and study
in detail the consequences for the observables.

We choose $n$, ${\rm{\bf v}}$ and $T$ as the independent variables.
We find that the  NR hydro equations with source terms are solved by the
following self-similar, ellipsoidally symmetric density, temperature and flow
profiles:
\begin{eqnarray}
n(t,{\rm {\bf r}^\p}) & = & n_0 {\frac{V_0} {V}}
		\mathrm{exp}\left({ -{\frac{r_x^{\p 2} }{2X^2}}
                	 -{\frac{r_y^{\p 2}}{2Y^2}}
  			-{\frac{r_z^{\p 2}}{2Z^2}} }\right),  \label{n} \\
{\rm {\bf v}^\p}(t,{\rm {\bf r}^\p}) &=&
    \left({\frac{\dot{X}}{X}}\, r_x^\p, \,\,
     {\frac{\dot{Y}}{Y}}\, r_y^\p, \,\,
     {\frac{\dot{Z}}{Z}}\, r_z^\p \right),  \label{v} \\
T(t) & = & T_i(t) 
	\left(  {\frac{\displaystyle\phantom{|}V_0}{\displaystyle%
	\phantom{|}V}} \right)^{1/\kappa},  \label{T} \\
T_i(t) & = & \kappa T_0 [1 + j_v(t)] 
	{\rm \exp}{\left( \int_{t_1}^{t} j_E(u) du \right)} \label{e:Tit}
\end{eqnarray}
where the variables are defined in a center of mass frame $K^\p$,
but with the axes pointing to the principial directions of the
expansion. This is the frame corresponding to the
System of Ellipsoidal Expansion (SEE), introduced
in ref.~\cite{ellobs}, where the relationship between the observables
in SEE and in the CMS of the fireball were analyzed in great
details.

The time dependent scale parameters are denoted by
$(X,Y,Z)$ the  typical volume of the expanding
system is $V = XYZ$, and the initial temperature and volume are
$T_0=T(t_0)$ and $V_0=V(t_0)$, and $n_0$ is a constant. The time
evolution of the radius parameters $X$, $Y$ and $Z$ is equivalent
to the classical motion of a particle in a non-central, time dependent
potential:
\bea
	\ddot X X = \ddot Y Y = \ddot Z Z =
	\frac{T_i(t)}{m} 
	\left( {\frac{\displaystyle\phantom{|}V_0}{\displaystyle%
	\phantom{|}V}} \right)^{1/\kappa},  \label{e:motion}
\eea
which correspond to a Hamiltonian motion of a mass point at $(X,Y,Z)$
driven by the following  non-central,  time dependent potential:
\bea
	H & = & \frac{P_x^2 + P_y^2 + P_z^2}{2 m} +
	T_i(t)
	\left( {\frac{\displaystyle\phantom{|}V_0}{\displaystyle%
	\phantom{|}V}} \right)^{1/\kappa},  
		\label{e:Ham}
\eea
	where $(P_x,P_y, P_z) = m(\dot X, \dot Y, \dot Z)$.
	See refs.~\cite{ellsol,cssol} for a similar problem,
	the solution of non-relativistic fireball hydrodynamics
	for self-similarly expanding, ellipsoidally symmetric case
	without source terms in the energy and the Euler equations.

	In all the cases, the normalization constant of the
	temperature increases in time
	corresponding to {\it reheating during inflation},
	given by eq.~(\ref{e:Tit}). 
	This is due to the external
	sources in the energy and momentum balance equations.
	This effect competes with a decreasing factor,
	governed by the expansion of the system, the rate of which is 
	controlled by the parameter $\kappa$. Note that
	for relativistic, massless particles, the speed
	of sound is simply $c_s^2 = 1/\kappa $.

	From this Hamiltonian formulation, 
	the effects of the equation of state and
	the effects from the source terms in 
	the energy and momentum balance equations
	can be qualitatively analyzed, and numerical solutions
	are also very easily obtained. 
	Softening the equation of state decreases slope of
	the repulsive potential in the corresponding mechanical 
	problem. As the coordinates
	$(X, Y, Z)$ correspond to the characteristic scales 
	of the expanding ellipsoids in the
	hydrodynamical system, softening the equation of state 
	means slowing down the expansion. 
	It is very interesting to observe the phenomena of 
	a ``stall" during a first order rehadronization 
	transition of a quark-gluon 
	plasma~\cite{gyulassy-dirk,shuryak,biro1,biro2}. 
	In our case, the acceleration of the fluid
	does not stop even for the softest equations of state,
	$1/\kappa \rightarrow 0$, 
	although the rate of expansion slows down substantially, 
	according to eq.~(\ref{e:motion}).

\section{Inflation generates strong transversal flows\label{flow}}

	From the analysis of the solution of fireball
	hydrodynamics with inflation, it follows that three 
	different effects may lead to the speeding up
	of the transverse expansion:

	{\it i)} Hardening the equation of state (corresponding to
	a decrease in the value of $\kappa$).
	This does not change the type of the expansion, 
	the coordinates $(X,Y,Z)$ increase as power-law type of
	functions of the time, as the slope of the
	potential term in eq.~(\ref{e:Ham}) is increased. 
	However, decreasing $\kappa$ makes
	the algebraic coefficients of the expansion larger.

	{\it ii)} Momentum transfer. A positive value of the
	factor $j_v(t)$ is also effective in increasing the
	height of the potential, however, the effect is only
	linear in $j_v$.

	{\it iii)} Energy transfer. This is the most effective method
	to inflate the fireball, as the increase in the volume changes
	its form from a characteristic power-law type of expansion
	to an exponential shape. Similar effect was discussed 
	for the case of spherically symmetric fireballs
	in ref.~\cite{cspeter}.

\section{Inflation, the gluon wind
 and scaling of HBT radii}\label{summ}

The observables: the shape, the slope parameters of the
single particle spectra, the various flow coefficients:
first, second and third flows, as well as the various
HBT radius parameters were calculated analytically
in ref.~\cite{ellobs} for the case when the source
terms were all set to zero.  We found that the introduction
of the source terms to the energy and the Euler equation
does not change the form of the solution,
only the time evolution of the source
parameters is modified. 
Hence, all the results of ref.~\cite{ellobs} 
are generalized to the case of inflating expansions.
We present some  numerical examples of the time evolution of the
 observables, using equations (8-37) of
ref.~\cite{ellobs}, that we do not recapitulate here. 
Although the steps are trivial, the results are 
literally surprising.

\subsection{Inflating hydro solutions with a gluon wind}
In this subsection we present the analysis of the expansion for 
the initial case of a heavy ion reaction, when energy and momentum
transfer to the soft modes is supposedly present, as a consequence
of the penetration of the highly energetic, perturbative gluons 
through the fireball. 
For simplicity, we assume that the momentum transfer is negligibly small,
 $j_v(t) = 0$. We assume that the QCD jet-quenching processes
and the related energy transport to the soft momentum modes
can be characterized by an approximately time independent 
coefficient $j_E(t) = j_0$, valid in a finite time
interval $(t_1,t_2)$.  In principle, the functions
$j_v(t)$ and $j_E(t)$ should be determined from QCD based
Monte-Carlo simulations, if a realistic simulation is attempted.
Here our aim was not yet a detailed fitting of the data, but the
qualitative understanding of the behaviour of measurables,
hence we analyzed the behaviour of the observables
for these reactions.
As high pt gluons are emitted from the beginning of the high energy
heavy ion collisions, it is convenient to choose $t_1 = 0$. The
end of the hard collisions can be estimated by the maximum of the
penetration time, $t_2 \approx 1.12 A^{(1/3)} / \gamma$. 
For clarity, we investigate analytically
the case when the fireball is initially spherical,
$X=Y=Z=R$ .

From eq.~(\ref{e:Ham}), we find that the radius of the fireball
inflates exponentially, and the coefficient of the temperature
also increases exponentially,
\bea
	R(t) & = & R_0 \exp\left( \frac{\kappa}{3 + 2 \kappa} j_0 t\right),\\
	T_i(t) & = & T_{i,0} \exp( j_0 t),\\
	T(t) & = & T_i(t) \left(\frac{R_0}{R}\right)^{3/\kappa} \, 
	= \, T_{i,0} \exp\left( \frac{2 \kappa}{3 + 2 \kappa} j_0 t\right).
\eea
Hence the local temperature $T(t)$ increases exponentially for all
$\kappa > 0$ values, because the exponentially increasing volume term
cannot compensate the even faster,  exponential
increase of the $T_i(t)$ factor. This effect is referred to as
reheating during the inflation, or reheating caused by the 
gluon wind.

Let us consider how such a scenario effects the experimentally
measurable slope parameters, $T_{\rm eff}$ and the HBT radius 
parameters, $R_{\rm eff}$. Following the lines of refs.~\cite{ellobs,cspeter},
these observables are given as
\bea
	T_{\rm eff} & = &  T + m \dot R^2 \, = \,
	( T_0 + \frac{\kappa}{3 + 2\kappa} m R_0^2)
	\exp( \frac{2 \kappa}{3 + 2 \kappa} j_0 t),\\
	R_{\rm eff} & \simeq & \frac{1}{j_0} 
	\left(\frac{ \kappa}{3 + 2 \kappa}\right)\sqrt{\frac{T_0}{m}}
	\exp\left(- \frac{\kappa}{3 + 2 \kappa} j_0 t\right),
\eea
which implies that the effective slope parameters
increase, while the effective homogeneity 
regions decrease exponentially with increasing time.
These features of the inflatory period remain qualitatively
similar for non-spherical initial conditions too,  
as shown in Figs. 1 and 2.

Hence inflation by the gluon wind may lead to the 
development of strong transverse flows as well as
small HBT radius parameters. However, after the
penetration of the nuclei through each other,
the production mechanism for high momentum gluons
comes to an end, and this implies that the
inflation by the gluon wind will be over. After this
time, the expansion will be driven by the hydrodynamical
equations without source terms.

Some numerical examples for asymmetric initial conditions
are shown on Fig. 1 and Fig. 2.
\begin{figure}[tbp]
\vspace{12.cm}
\begin{center}
\includegraphics{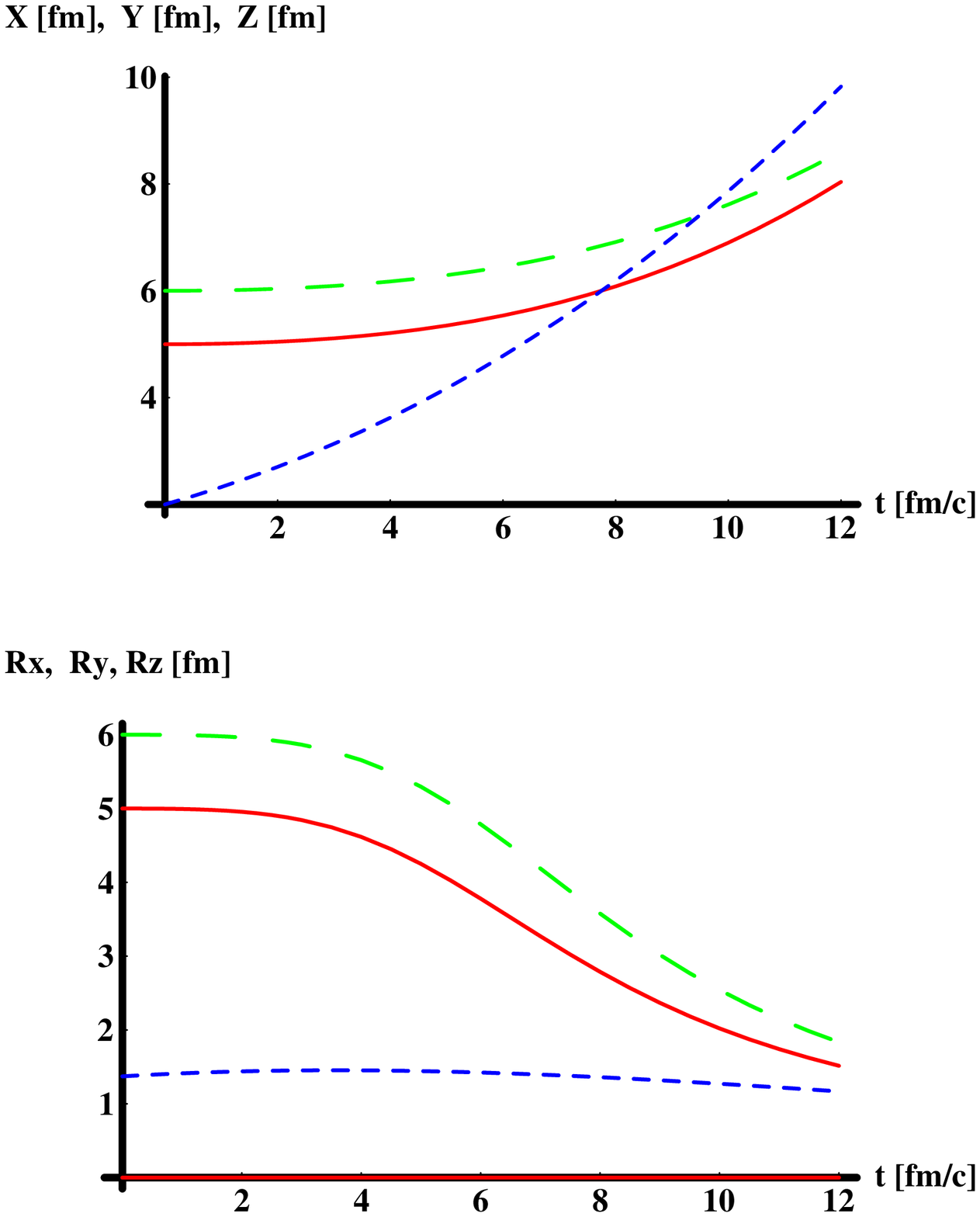}
\end{center}
\vspace{-2.5cm}
\caption{The effect of the gluon wind on the hydro
solution.  We plot the evolution of $(X,Y,Z)$,
the principal axis of the expanding ellipsoid,
and the corresponding measurable HBT radius parameters,
$(R_x, R_y, R_Z)$ for expanding fireballs in hydrodynamics, 
with inflation and source terms in the energy equation, due to a 
gluon wind. 
 The characteristic geometrical sizes increase exponentially
with time, while the measurable HBT radii decrease exponentially with time.
The characteristic time scales to reach the equality for all 
the HBT radii become much shorter than on Fig. 3, which is the
case without inflation.
The initial conditions are $(X_0,Y_0,Z_0) = (5,6,2)$ fm,
$(\dot X, \dot Y, \dot Z) = (0.,0.,0.3)$, $m = 940$ MeV,
$T_{i,0} = 94$ MeV and $j_E = j_0 = 0.3 $ c/fm.
}
\end{figure}

\begin{figure}[tbp]
\vspace{12.cm}
\begin{center}
\includegraphics{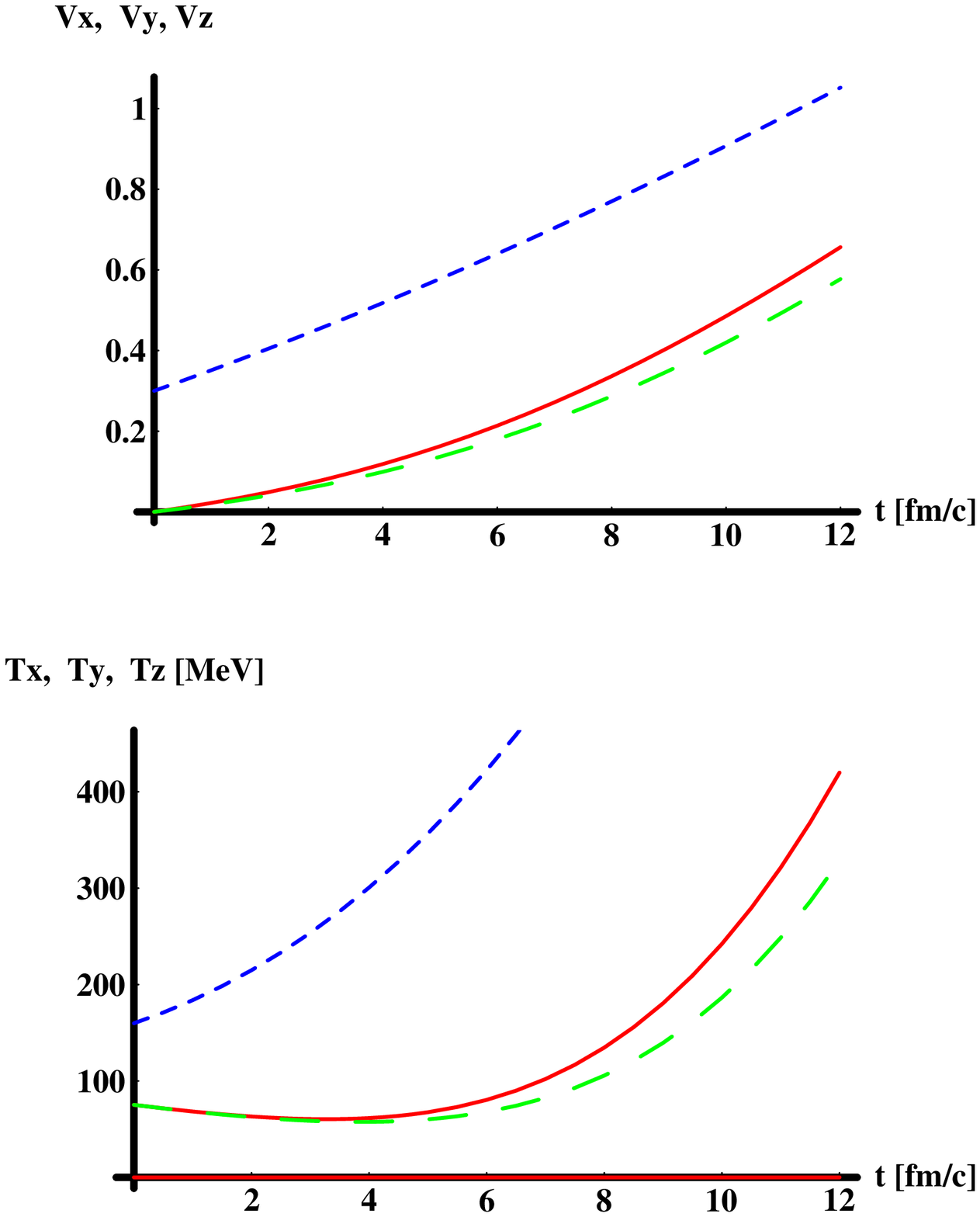}
\end{center}
\vspace{-2.5cm}
\caption{The effect of the gluon wind on the
observables  for the inflating fireballs, the same solution
as on Fig.1.  
 During
the period of the inflation, neither the slopes nor the
HBT radii saturate. 
}
\end{figure}
 
\subsection{Asymptotic hydro without gluon wind effects}

Let us  observe, that the analytic results indicate a fast transverse
flow at the end of the inflatory period.

An asymptotic solution of the hydrodynamical  problem is
given by the large time, large volume approximation, which implies
\bea
	X\ddot X & = &  Y \ddot Y  = Z \ddot Z \rightarrow 0,\\
	\dot X & \rightarrow & \dot X_a, \quad \quad\! 
	X \rightarrow \dot X_a t,\\
	\dot Y & \rightarrow & \dot Y_a, \quad \quad\, 
	Y \rightarrow \dot Y_a t,\\
	\dot Z & \rightarrow & \dot Z_a, \quad \quad\, 
	Z \rightarrow \dot Z_a t, \\
	T & \rightarrow& T_a (t_a/t)^2.
\eea
	Substituting this solution into the expressions for the
	observables, we obtain a very interesting result,
	namely that the observables freeze-out
	in the sense that their value becomes independent from
	the time of the observation in the asymptotically late
	period of the expansion, when the inflation effects are 
	already over.
\bea
	T_x &\rightarrow  & m \dot X_a^2, \\
	T_y &\rightarrow & m \dot Y_a^2,\\
	T_x &\rightarrow & m \dot Z_a^2,\\
	R_{\rm side} & \approx & R_{\rm out} \, \approx \, R_{\rm long}
	\rightarrow t_a \sqrt{T_a/m} ,
\eea
	independently of the time, for asymptotically large values of the time,
	which implies that the region of homogeneity becomes spherically
	symmetric. Due to this reason, it is possible to show, that
	the cross-terms all vanish
\be
	R_{s,o}^2 \approx R_{o,l}^2 \approx R_{s,l}^2 \rightarrow 0
\ee
	in the late stage of the expansion, not only in the system
	of ellipsoidal expansion~\cite{ellobs},
	but also in the frame of the observation.

	Hence the observable part of the expanding fireball becomes
	spherically symmetric, after the inflation to large source sizes
	all the initial ellipsoidal geometrical symmetry is washed out
	from the local regions of homogeneity. 
	Only the thermal scales~\cite{3d,qm95,cs-review,ellobs}
	dominate the observable HBT radii, and these scales become 
	- due to the very nature of the thermal smearing - 
	direction independent.

\begin{figure}[tbp]
\vspace*{12.0cm}
\begin{center}
\includegraphics{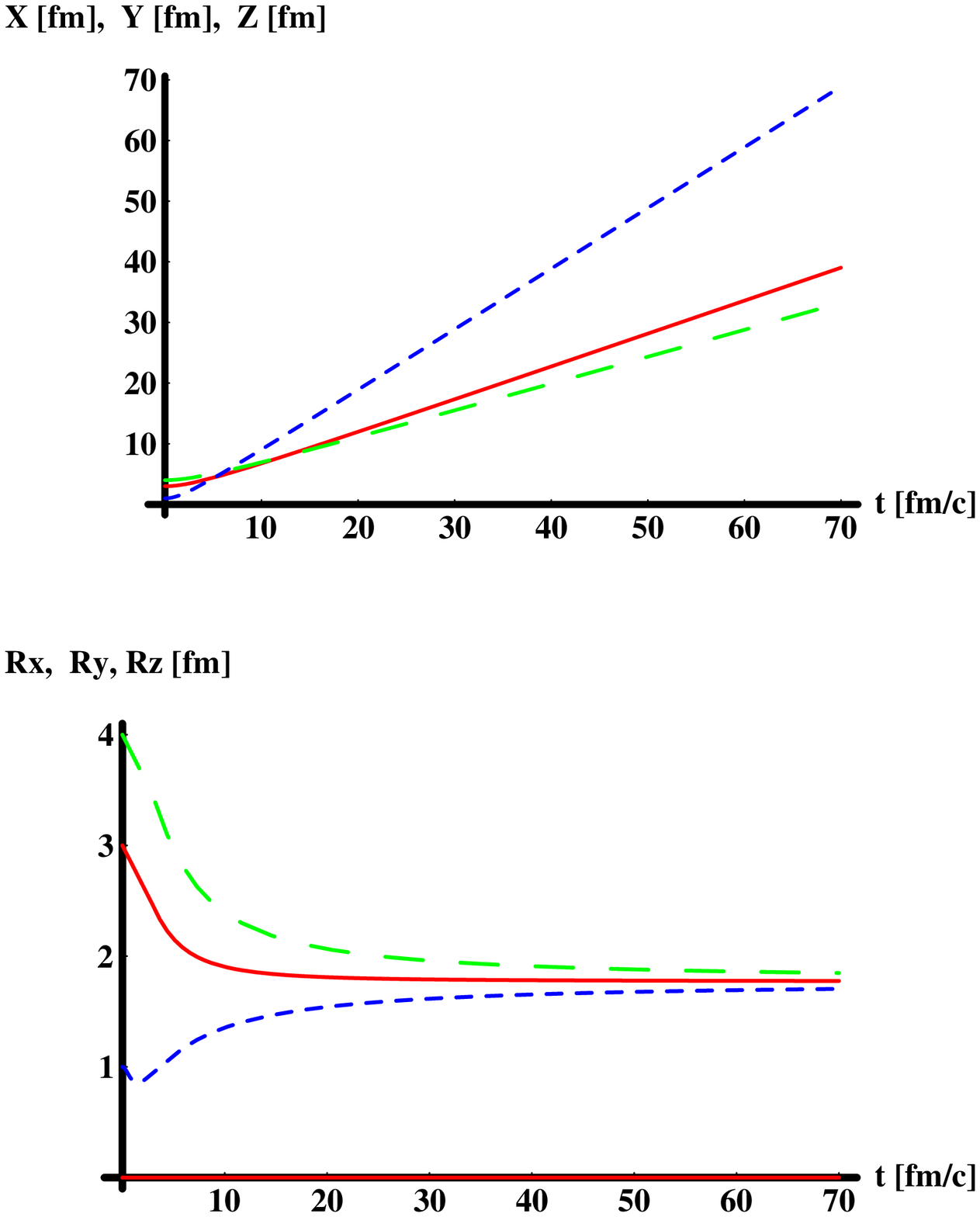}
\end{center}
\vspace{-2.5cm}
\caption{
The time evolution of $(X,Y,Z)$,
the principal axis of the expanding ellipsoid,
and the corresponding measurable HBT radius parameters,
$(R_x, R_y, R_Z)$ for expanding fireballs in hydrodynamics, 
without inflation and source terms. 
The initial conditions are $(X_0, Y_0, Z_0) = (6,5,2)$ fm,
$(\dot X_0,\dot Y_0, \dot Z_0) = (0,0,0)$, $T_i = 470$ MeV 
$m = 940 $ MeV. The acceleration period
is over after about 10 fm/c, the solution tends to a coasting, 
ellipsoidal fireball. The HBT radii approach the same, constant
and direction independent value, although the actual source sizes
keep on expanding and remain direction dependent. Note the qualitative
similarity to the results of ref.~\cite{Humanic}.
}
\end{figure}

\begin{figure}[tbp]
\vspace{12.cm}
\begin{center}
\includegraphics{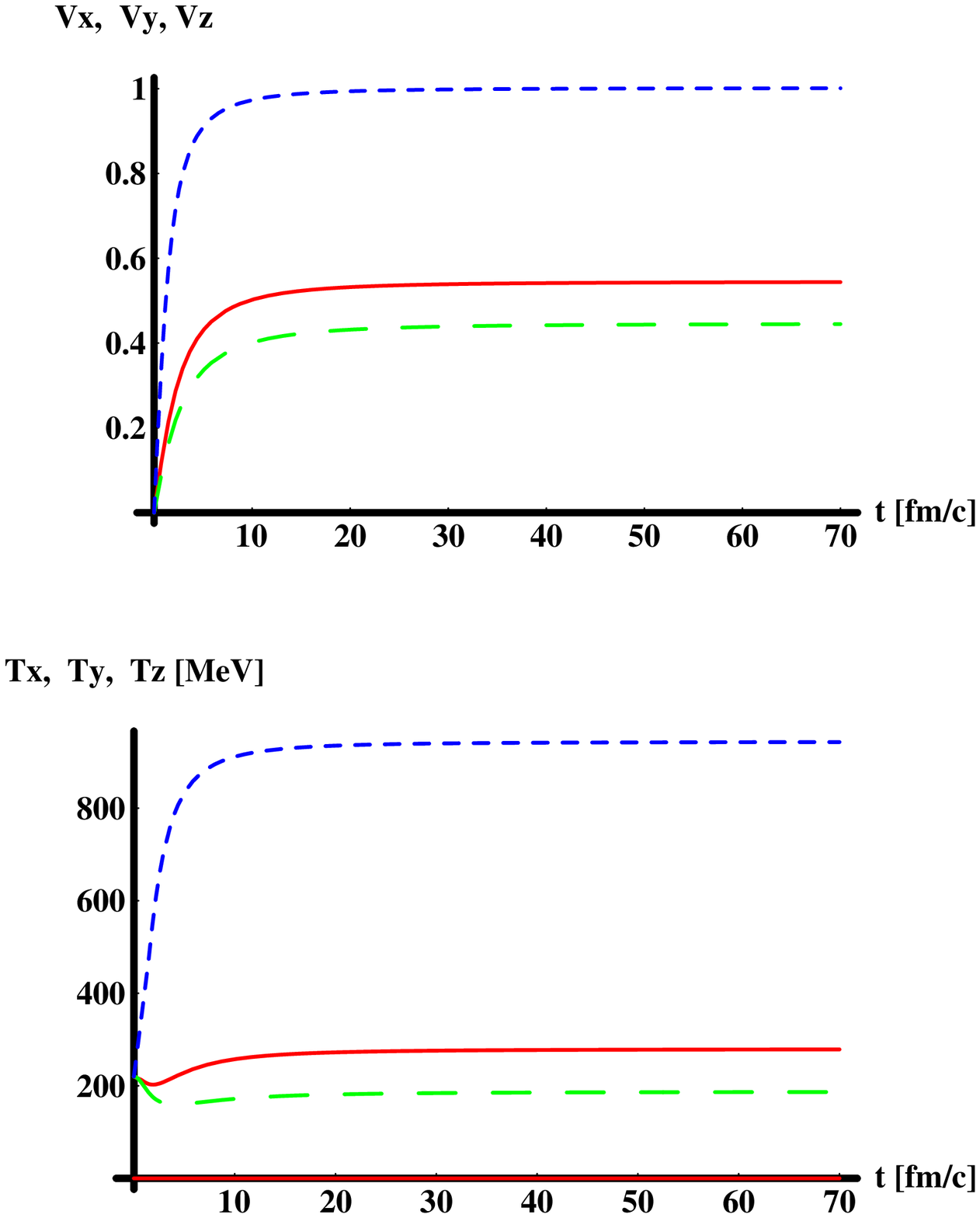}
\end{center}
\vspace{-2.5cm}
\caption{The time evolution of the Hubble constants
$(\dot X, \dot Y, \dot Z)$ and the measurable  
slope parameters $(T_x, T_y, T_z)$ for the hydro solution shown
in Fig. 3.  The saturation of the slopes of the single particle spectra
happens at the same time as that of the Hubble
constants, which happens earlier than the saturation of
the measurable HBT radius parameters.
}
\end{figure}

\subsection{Relativistic corrections}

	We find that the flow field for asymptotically large
	times approaches a spherically symmetric, Bjorken type
	flow field, $u^\mu = x^\mu/\tau = \gamma (1, {\bf  v})$
	with ${\bf v} = {\bf r}/t$. This is the same flow profile
	as the Hubble flow of our Universe. Furthermore,
	due to the spherical symmetry of the asymptotic flow 
	profile, the local lengths of homogeneity will also
	become spherically symmetric. Due to the large value of the
	geometrical scales, the lengths of homogeneity will
	measure only the {\it thermal} 
	scales~\cite{3d,qm95,ellobs,cs-review} ,
	and due to the homogeneity of the temperature and the
	asymptotic flow profile, the volume of homogeneity
	becomes spherical, and all the thermal length scales will
	be given by 
	$R_{out} \simeq R_{side} \simeq R_{long} 
	\simeq t \sqrt{\frac{T(t)}{m}}$. Interestingly,
	this thermal length scale will become a constant of
	motion in the late stages of the expansion, as the temperature
	decreases in time as $T(t) = T_f (t_f/t)^2$, so all the
	radius parameters tend to $t_f \sqrt{\frac{T_f}{m}}$.

	The relativistic corrections to this result can be
	estimated following the lines of reasoning and the
	conditions given in ref.~\cite{3d,qm95} when analyzing
	the connection between the observables and a relativistic
	generalization of the source functions. Under certain
	conditions, spelled out in great details in ref.~\cite{3d}
	and investigated numerically also in ref.~\cite{3d-test},
	the relativistic form of  result is
\begin{equation}
	R_{out} \simeq R_{side} \simeq R_{long}  \simeq
	\tau_f  \sqrt{\frac{T_f}{m_t}},
\end{equation}
	where $\tau_f= \sqrt{t^2 - r_z^2}$ at the time when the
	accelerationless expansion sets in, 
	$m_t = \sqrt{m^2 + p_x^2 + p_z^2}$ is the transverse
	mass  and $T_f$ is the value of the local temperature
	in the center of the transverse plane at the propertime $\tau_f$.

\subsection{Microscopic dynamics}
	It is interesting to note, that the time evolution of the
	observables from the ellipsoidally symmetric hydro solution
	is very much similar to the time evolution given by T. Humanic's
	cascade code, where the pions, kaons, nucleons and other hadronic
	resonances take part in rescattering, starting from a very
	hot and dense initial state~\cite{Humanic}.
	The realization that the HBT radii may reach a saturation limit,
	but somewhat later, than the single particle spectra, was
	recently predicted in ref.~\cite{Dumitru}, and confirmed in
	Figs. 3, 4. Strong transverse
	push was generated also by massive quark matter produced
	in the middle of Au+Au collisions from a cascade model in
	ref.~\cite{Lin} . 
	It seems that a various gases of massive quanta 
	that undergoes strong rescattering yields qualitatively similar
	results to the hydrodynamic solution that we described here.


\section{Summary and Conclusions}\label{concl}

	We find that an inflatory period automatically leads to a spherical
	symmetry for the observable HBT radius parameters in an analytically
	solvable model of fireball hydrodynamics with source terms motivated
	by the flux of high-pt gluon wind through the expanding hadronic
	matter, created in high energy collisions of heavy ions.
	The effect is very similar in nature how a flat and spatially
	homogeneous Universe is obtained after a period of inflation in
	astrophysics. Our results underline the similarity between
	the physics of 'Little Bangs' at the smallest experimentally
	accessible scales and the physics of the 'Big Bang' of our Universe,
	at the largest observable scales.

	We also find, that the 
	form of the observables is {\it exactly} the same
	for cases with or without inflation, and for various
	changes of the equation of state in the intermediate
	steps of the time evolution.  This implies that various 
	different equations of state from various different initial 
	conditions may lead to {\it exactly} the same hadronic
	final state. 
	Furthermore, the time evolution of the parameters
	of the hydrodynamical solution is sensitive not only 
	to the equations of state but also to the magnitude of 
	the source terms in the energy and momentum balance equations, 
	which may provide non-equilibrium mechanisms to connect 
	the initial state to a violently exploding final state.

	These results imply that {\it i)} the hadronic final
	state of fireball evolution does not remember the path
	(hence any earlier transient phase of matter) of its time
	evolution, but {\it ii)} given the equation of state,
	and given the possible source terms in the energy and the
	Euler equation, the initial state can still be reconstructed,
	from the observables in the final hadronic state.

	Further studies are in progress in for the publication
	of the ellipsoidally symmetric solutions of relativistic
	hydrodynamics and for the calculation of the observables
	corresponding to these relativistic generalizations
	of the present results.

{\it Acknowledgments:} This study has been supported by 
the OTKA grants T034262, T038406,  by the US - Hungarian NSF - OTKA - MTA 
grant 0089462 and by an Alumni Initiatives Award of the
Fulbright Foundation.

\vfill\eject 
\end{document}